\baselineskip=24pt
\magnification=1200
\hsize=6.5truein
\vsize=9truein

\def\h{{1\over 2}}
\hyphenation{Di-rac}
\hyphenation{fer-mi-ons}
\line{\hfill UdeM-LPS-TH-93-144, hep-th/9304078}
\centerline{\bf Violations of the String Hypothesis}
\centerline{\bf in the}
\centerline{\bf Solutions of the Bethe Ansatz Equations in the XXX-Heisenberg
Model}
\vfill
\vskip.5truecm
\centerline{Karl Isler$^*$}
\centerline{Instituut voor Theoretische Fysica}
\centerline{Rijksuniversiteit Utrecht}
\centerline{Princetonplein 5}
\centerline{Utrecht, Netherlands}
\centerline{and}
\centerline {M. B. Paranjape,}
\centerline{Institut f\"ur Theoretische Physik,}
\centerline{Universit\"at Innsbruck, Technikerstrasse 25, Innsbruck, Austria,
A-6020}
\centerline{ and}
\centerline{ Laboratoire de physique nucl\'eaire$^\dagger$, D\'epartement de
physique} \centerline{Universit\'e de Montr\'eal, C. P. 6128 succ. ``A"}
\centerline{Montr\'eal, Qu\'ebec, Canada, H3C 3J7}
\vskip1truecm
\noindent
\centerline{\bf Abstract}
We study the equations for the quasi-momenta which characterize the
wave-functions in the Bethe ansatz for the XXX-Heisenberg model.  We show in
a simple analytical fashion,
that the usual ``string hypothesis" incorrectly predicts the
number of real solutions and the number of complex solutions for $N>21$ in
the sector with two spins flipped, confirming the
work of Essler et al.  Two complex pair solutions
drop out and form two additional real pair solutions.  We also introduce a new
set of variables which allows the equations to be written as a single
polynomial equation in one variable.  We consider in some detail the
case of three spins flipped.
\vskip1truecm \noindent
The XXX-Heisenberg model
appears to be fundamental to many of the recent interesting
developments in modern quantum field theory and mathematical physics.
Integrable
models, conformal field theories and quantum groups are seen to arise from
various
limits, extensions and generalizations$^1$.  It is, in fact, the original model
considered by Bethe$^2$, which gave rise to the celebrated ``Bethe ansatz"
solutions.  Recently, there appeared an article$^3$, studying the Bethe ansatz
equations in the two particle sector of the spin one-half model.  It was shown
that, as the number of lattice sites, $N$, increases past 21, two new
real pair solutions appear.  These solutions do not fit into the
conventional scheme of classification of the solutions to the Bethe ansatz
equations which causes some anxiety as to the verity of the completeness of the
full set of $SU(2)$ extended Bethe ansatz states.  The full set of states, just
by
counting, should number $2^N$.   This, however, is not a problem.  As the real
pair
solutions appear, complex pair solutions disappear, conserving the total number
of states.  We analyze the Bethe ansatz equations analytically and confirm
their result; two complex pair solutions disappear
simultaneously giving two new real pair solutions.
\vskip.5truecm\par\noindent
The spin $\h$
XXX-Heisenberg model with $N$ spins corresponds to the Hamiltonian $$
H={J\over 4}\sum_{i=1}^N\left(\vec\sigma_i\cdot\vec\sigma_{i+1}
-1\right)\eqno(1)
$$
with $\vec\sigma_{N+1}=\vec\sigma_1$, and the $\vec\sigma_i$ are just the Pauli
matrices for each $i$.  The Bethe ansatz in the $M$ particle sector corresponds
to eigenfunctions of the form
$$
|\Psi\rangle =\sum_{x_1<x_2<\cdots <x_M} \psi(x_1,\cdots ,x_M)\sigma^-_{x_1}
\sigma^-_{x_2}\sigma^-_{x_3}\cdots\sigma^-_{x_M}|0\rangle\eqno(2)
$$
with
$$
\psi(x_1,\cdots ,x_M)=\sum_{P\in\, S_M}e^{\{ i\sum_{j=1}^Mk_{P(j)}x_j +
i\sum_{j<l\atop P(j)>P(l)}\phi_{P(j),P(l)}\} },\eqno(3)
$$
$|0\rangle$ is the ferromagnetic state with all spins up, $P$ is a permutation
on $M$ objects and $\phi_{i,j}$ satisfies
$$
2\cot ({\phi_{i,j}\over 2})=\cot ({k_{i}\over 2})-\cot ({k_{j}\over
2}).\eqno(4)
$$
$k_1\cdots k_M \in [0,2\pi]$ are the quasi-momenta or spectral parameters.  The
wave-function is symmetric under interchange of any two coordinates $x_j$.
\vskip.5truecm\par\noindent
Periodic boundary conditions
$$
\psi(x_1,\cdots ,x_{M-1},N+1)=\psi(1,x_1,\cdots ,x_{M-1})\eqno(5)
$$
imply the $M$ coupled equations
$$
e^{ik_jN}=\prod_{l=1\atop l\ne j}^Me^{i\phi_{j,l}}\quad\quad\quad
j=1,2,\cdots ,M.\eqno(6)
$$
\vskip.5truecm\par\noindent
The usual set of variables considered are $\Lambda_j=\cot ({k_{j}\over 2})$
which give the equivalent form for the equations $$
\left(e(\Lambda_j)\right)^N=\prod_{l=1\atop l\ne
j}^Me\left({\Lambda_j-\Lambda_l
\over 2}\right)\quad\quad\quad j=1,2,\cdots ,M,\eqno(7)
$$
with
$$
e(\Lambda )={\Lambda +i\over\Lambda -i}.\eqno(8)
$$
The energy of the state is then given by
$$
E(\Lambda_1,\cdots ,\Lambda_M)=\sum_{j=1}^M{-2J\over \Lambda^2_j+1}.\eqno(9)
$$
\vskip.5truecm\par\noindent
In the large $N$ limit, the ``string hypothesis"$^4$ states that, for fixed
$M$,
any solution $\Lambda_1,\cdots ,\Lambda_M$ consists of ``strings" of the form
$$
\Lambda_\alpha^{n,j}=\Lambda_\alpha^n+i(n+1-2j)+o(e^{-\delta N})\quad\quad\quad
j=1,\cdots ,n\eqno(10)
$$
where $n\ge 1$ gives the length of the string, $\alpha $ labels strings of a
given length, $j$ specifies the imaginary part of $\Lambda $ and $\delta > 0$.
With such a hypothesis, the Bethe ansatz equations only involve the real parts,
$\Lambda_\alpha^n$.  The
solutions are parametrized by (half-odd) integer numbers $I_\alpha^n$ for
$N-M_n$ (even) odd, where $M_n$ is the total number of strings of length
$n$.  Clearly $\sum_{n=1}^\infty nM_n=M$.  It is generally believed that there
is
a $1-1$ correspondence between solutions of the Bethe ansatz equations and sets
of
independent, non-repeating integers $I_\alpha^n$, ($I_\alpha^n\ne I_\beta^n$
for
$\alpha\ne\beta$, i.e., no two strings of the same length contain the same
integer, within one solution set of $\Lambda_1\cdots\Lambda_M$)$^4$.
\vskip.5truecm\par\noindent
The Bethe ansatz equations for $M=2$ are
$$ \eqalign{
\left(e(\Lambda_1)\right)^N&=e({\Lambda_1-\Lambda_2\over 2})\cr
\left(e(\Lambda_2)\right)^N&=e({\Lambda_2-\Lambda_1\over 2}).}\eqno(11)
$$
This set of equations has the following symmetries,
$\Lambda_i\rightarrow\Lambda_i^*$,  $\Lambda_i\rightarrow -\Lambda_i$
and $\Lambda_1\leftrightarrow\Lambda_2$.  Now
$$
e({\Lambda_1-\Lambda_2\over 2})={{\Lambda_1-\Lambda_2\over 2}+i\over
{\Lambda_1-\Lambda_2\over 2}-i}={\Lambda_1+i-(\Lambda_2-i)\over
\Lambda_1-i-(\Lambda_2+i)}.\eqno(12)
$$
Replacing for $\Lambda_i$ with $\Lambda_i={i(e(\Lambda_i)+1)\over
e(\Lambda_i)-1}$ gives
$$
e({\Lambda_1-\Lambda_2\over
2})=-\left({e(\Lambda_1)e(\Lambda_2)-2e(\Lambda_1)+1\over
e(\Lambda_1)e(\Lambda_2)-2e(\Lambda_2)+1}\right).\eqno(13)
$$
We choose to work with the variables $X_i=e(\Lambda_i)$.  In terms of these
variables, real solutions (in the $\Lambda_i$) are mapped to the unit circle
while complex conjugate pairs are mapped to complex pairs which are related
by, $(z,{1\over z^*})$. Then we get
$$
X_1^N=
-\left({X_1X_2-2X_1+1\over X_1X_2-2X_2+1}\right)\quad{\rm and}\quad X_2^N=
-\left({X_1X_2-2X_2+1\over X_1X_2-2X_1+1}\right). \eqno(14a,b)
$$
Multiplying these equations
together gives, $$
(X_1X_2)^N=1,\quad{\rm hence}\quad X_1X_2=\omega\eqno(15)
$$
where $\omega$ is an $N$th root of unity.  Thus replacing
$X_2={\omega\over{X_1}}$ in (14a), yields
$$
X_1^N=-\left({\omega -2X_1+1\over \omega -2{\omega\over X_1}+1}\right)=
-X_1\left({\omega -2X_1+1\over \omega X_1 -2\omega +X_1}\right).\eqno(16)
$$
Assuming $X_1\ne{2\omega\over\omega+1}$, or $0$, we get
$$
X_1^{(N-1)}((\omega +1)X_1-2\omega) + \omega -2X_1+1=0\eqno(17)
$$
ie.
$$
(\omega +1)X_1^N-2\omega X_1^{(N-1)}-2X_1+(\omega +1)=0.\eqno(18)
$$
We will be interested in the real roots in
terms of $\Lambda_1$. These are mapped to roots on the unit circle in terms of
$X_1$.  Furthermore
$$ \Big|{2\omega\over (\omega +1)}\Big|^2={4\over (\omega
+1)(\omega^* +1)}= {4\over 2+2{\rm Re}\omega}={2\over 1+{\rm Re}\omega}>1
\eqno(19)
$$ for
$\omega\ne 1$.  Thus the denominator by which we multiplied may only vanish for
$\omega = 1$ and $X_1=1$, which gives $X_2=1$.  The corresponding wave function
vanishes identically as the Bethe ansatz wave functions respect the Pauli
principle. We want the roots of equation (18), for each choice of
$\omega$.  This gives $N$ polynomials, each of order $N$, which is a total of
$N^2$ roots.  $X_1=\sqrt\omega$, however, is always a solution
with $X_2=\sqrt\omega$.  These wave functions also vanish because of the Pauli
principle.  Furthermore, if $x$ is a root, then $\omega\over
x$ is also a root.  Each one gives the same Bethe ansatz wave
function, only the roles of $X_1$ and $X_2$ are exchanged.  Therefore, we get a
total of ${N^2-N\over 2}=({N\atop 2})$ different wave functions.  This is
exactly
the dimension of the subspace of states with two spins flipped.  There are
exactly $({N\atop 2})$ independent ways of flipping two spins among $N$.
\vskip.5truecm\par\noindent
The equation is cast in a more symmetric
form with the replacement $X_1\rightarrow \sqrt\omega \tilde X_1$ and
correspondingly $Y_1\rightarrow \sqrt\omega \tilde Y_1$.  These satisfy the
symmetric relation $\tilde Y_1 ={1\over \tilde X_1}$.  We first consider the
case $N$ odd, where it is always possible to take $\sqrt\omega^N=-1$.  This
means with $\omega =e^{i{2\pi m\over N}}$, for $m$ odd we take  $\sqrt\omega
=e^{i{\pi m\over N}}$ but for $m$ even we must take $\sqrt\omega =e^{-i{\pi
(N-m)\over N}}$.  The set of $\{\sqrt\omega\}$ for $m$ odd are exactly the
inverses (complex conjugates) of the set for $m$ even, except for $\omega =1$,
which is not so paired.  Then taking into account $\sqrt\omega^N=-1$, we get
$$ \eqalign{ 0&=(\omega +1)\sqrt\omega^N\tilde
X_1^N-2\omega\sqrt\omega^{N-1} \tilde X_1^{(N-1)}-2\sqrt\omega\tilde
X_1+(\omega +1)\cr &=-(\omega +1)\tilde X_1^N+2\sqrt\omega \tilde X_1^{(N-1)}-
2\sqrt\omega \tilde X_1+(\omega +1)\cr
&=-\sqrt\omega\left((\sqrt\omega+{1\over\sqrt\omega})\tilde X_1^N-2\tilde
X_1^{N-1}+2\tilde X_1- (\sqrt\omega+{1\over\sqrt\omega})\right)}\eqno(20)
$$  which
yields
$$ \cos({{\theta\over 2}})\tilde X_1^N-\tilde X_1^{(N-1)}+\tilde X_1-
\cos({\theta\over 2}) =0,  \eqno(21)
$$
with the definition $\sqrt\omega
=e^{i({\theta\over 2})}$.  The symmetries of equation (11) now translate into
$(\tilde X_1\rightarrow{1\over\tilde X_1^*},\omega\rightarrow\omega)$, $(\tilde
X_1\rightarrow{1\over\tilde X_1},\omega\rightarrow\omega^*)$ and $(\tilde
X_1\rightarrow{1\over\tilde X_1},\omega\rightarrow\omega)$.
\vskip.5truecm\par\noindent
The roots of Equation (21) come in pairs, if $z$ is a
root, so is $1\over z$.   Furthermore, equation (21) has real coefficients,
thus
complex solutions come in complex conjugate pairs.  The natural grouping of
complex
solutions is actually in quartets, $(z,z^*,{1\over z},{1\over z^*})$.  Real
solutions in terms of the $\Lambda$ variables are mapped to solutions
on the unit circle, thus the quartets degenerate into pairs for these.
Complex solutions in terms of the $\Lambda$ variables are then mapped
off the unit circle, and should generally come in quartets.  Therefore, if
there
is truly only one complex pair in terms of the $\Lambda$ variables, it must be
mapped to a real pair, $r,{1\over r}$ in terms of the $\tilde X$ variables.
$\tilde X_1=r$ fixes the solution for the corresponding $\Lambda =
i{r\sqrt\omega
-1\over r\sqrt\omega +1}=-2{r\sin\theta\over r^2+1}+i{r^2-1\over r^2+1}$.   It
would
be interesting to find true quartet solutions. Below we only find pairs.
Equation
(21) is the same for $\sqrt\omega$ and for $1\over\sqrt\omega$.  Although the
equations, and consequently the solutions, are the same in the tilde variables,
the solutions in terms of the original $X$ variables are of course different,
giving rise to different  Bethe ansatz wave functions.
\vskip.5truecm\par\noindent
Now we look for solutions on the unit circle,
$\tilde X_1=e^{i\alpha}$.  Equation (21) becomes  $$ \cos({\theta\over
2})e^{iN\alpha}-e^{i(N-1)\alpha}+e^{i\alpha}-\cos({\theta\over 2}) =0,
\eqno(22)
$$  which simplifies magically to
$$
\eqalign{
0&=e^{iN\alpha\over 2}\cos({\theta\over
2})(e^{iN\alpha\over 2}- e^{-iN\alpha\over 2})- e^{i\alpha}e^{i(N-2)\alpha\over
2} (e^{i(N-2)\alpha\over 2}-e^{-i(N-2)\alpha\over 2})\cr &=e^{iN\alpha\over
2}2i\left(\cos({\theta\over 2})\sin ({N\alpha\over 2})- \sin ({(N-2)\alpha\over
2})\right).}\eqno(23)
$$
Thus
$$
\cos({\theta\over 2})={\sin ({(N-2)\alpha\over 2})\over\sin ({N\alpha\over
2})}.
\eqno(24)
$$
This equation is easily studied graphically, see figure (1).
$\alpha\in [0,2\pi ]$.  The right hand
side has zeros at $\alpha = {2k\pi\over N-2}$ for $k=1,2,3,\cdots ,N-3$ and
poles at $\alpha = {2k\pi\over N}$ for $k=1,2,3,\cdots ,N-1$.  The value at
$\alpha = 0$ is ${N-2\over N}$ but at $\alpha =\pi$ it is $-1$.  We
get a root for each intersection of a horizontal line with $y=
\cos({\theta\over
2})$ with the curve $y={\sin ({(N-2)\alpha\over2})\over\sin ({N\alpha\over
2})}$.
\input psbox.tex
\centinsert{\pscaption{
\boxit{\psboxscaled{800}{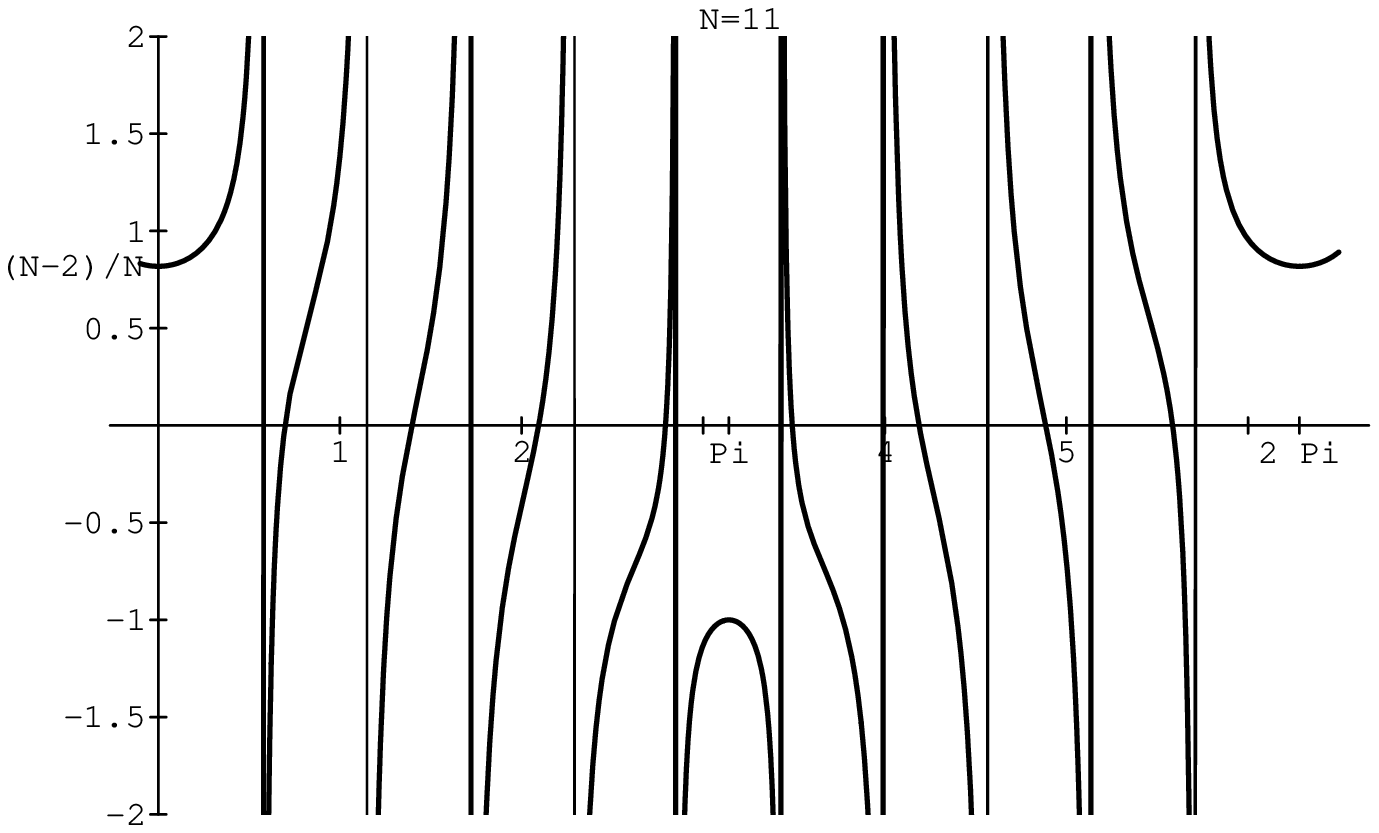}}}
{{\bf Figure 1}: $f(\alpha)={\sin ({(N-2)\alpha\over 2})\over\sin
({N\alpha\over 2})}$, $N=11$}}
\vskip.5truecm\par\noindent
By counting the intersections it is evident from figure (1), that there is a
root for each zero.  There are always $N-3$ such roots.  An additional pair of
real roots can appear if
$$
\cos({\theta\over 2})\ge{N-2\over N},\eqno(25)
$$
giving $N-1$ real roots, but  $\tilde X_1=1$, $(X_1=\sqrt\omega)$, is also a
root
of the original equation, giving totally $N$  roots.  This exhausts all roots
of
the polynomial and there are no complex conjugate pairs (for $\Lambda$).
Thus we find
$$
\cos({\theta_{\rm critical}\over 2})={N-2\over N}.\eqno(26)
$$
For large $N$, this will have solutions for ${\theta\over 2}$ near 0 or $2\pi$,
but because of reflection symmetry about $\alpha =\pi$ we need only search near
$\alpha = 0$.   Assuming $\theta$ is small and making an expansion in $\theta$
and $1\over N$, we get
$$
1-{1\over 2}({{{\theta_{\rm critical}\over 2}}})^2 +\cdots =1-{2\over
N}\eqno(27)
$$
giving
$$
\theta_{\rm critical}= {4\over\sqrt N}.\eqno(28)
$$
The values taken by ${\theta\over 2}$ (the condition that $\sqrt\omega^N=-1$
must be
satisfied) give
$$
{\theta\over 2} = {\pi m\over N}\quad {\rm for} \quad m=1,3,5,\cdots ,N-2,
\eqno(29)
$$
or
$$
{\theta\over 2}  = \pi-{\pi m\over N} \quad {\rm for} \quad m=2,4,6,\cdots
,N-1, \eqno(30)
$$
($N$ is odd).  Hence, for the odd series in $m$, we get
$$
m<{2\over\pi}\sqrt N\eqno(31)
$$
and for the even series in $m$,
$$
|N-m|<{2\over\pi}\sqrt N.\eqno(32)
$$
The number of new solutions behaves like $\sqrt N$.
\vskip.5truecm\par\noindent
The first $N$
for which we have a new solution, (to equation (25) actually), is $N=23$,
when $m=3$ is allowed, next at
$N=63$, $m=5$ is allowed and so on.   (The solution at $m=1$ is accounted for
in
the string hypothesis set of solutions.  It corresponds to the  action of the
lowering operator applied to the corresponding state in the $M=1$ sector,
in the context of the $SU(2)$ extended Bethe ansatz.)
\vskip.5truecm\par\noindent
Analyzing in more detail the complex pair of
roots, we insert $\tilde X_1=r$ into equation (21), giving
$$
\cos({\theta\over 2})= {r(r^{N-2}-1)\over r^N-1}=f(r).\eqno(33)
$$
\centinsert{\pscaption{
\boxit{\psboxscaled{800}{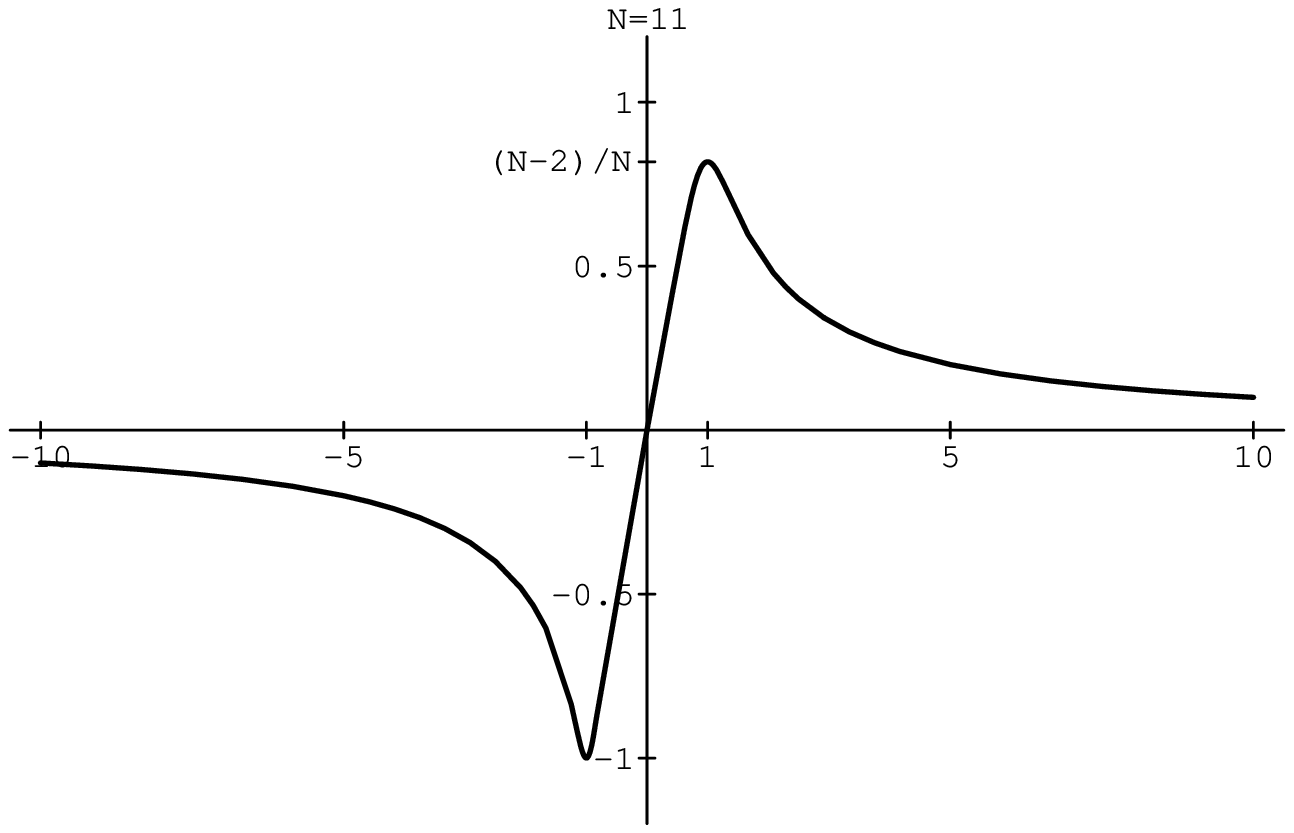}}}
{{\bf Figure 2}: $f(r)={r(r^{N-2}-1)\over r^N-1}$, $N=11$}}
\par\noindent
It is easy to see that $f(r)$ is monotone decreasing for $r>1$.  Therefore,  it
is
monotone increasing for $0<r<1$, as $f({1\over r})=f(r)$.  The maximum occurs
at $r=1$, $f(1)={N-2\over N}$.  Evidently there are two roots for each $0<
\cos({\theta\over 2})<{N-2\over N}$, as we can see from figure (2).  This is
exactly the same condition as equation (25), as expected.  We see that for
${\theta\over 2}$ near $\pi\over 2$ the solution $(r,{1\over r})$ is $(\infty
,0)$,
which implies $\Lambda\approx -2{r\sin\theta\over r^2+1}\pm i$, as
predicted by the string hypothesis.  For ${\theta\over 2}$ near zero, however,
the imaginary part becomes arbitrarily small and the string hypothesis is
grossly violated, finally to the extreme that the complex solutions drop out
all together.
\vskip.5truecm\par\noindent
For $r<0$,
$$
f(r)=-|r|({|r|^{N-2}+1\over |r|^N+1})\eqno(34)
$$
It is again easy to see that $f(r)$ decreases monotonically from $0$ to $-1$ as
$r$ varies from $-\infty$ to $-1$ after which it rises monotonically to $0$,
since
always $f({1\over r})=f(r)$.  Thus, we always get exactly one complex pair for
${\theta\over 2}\in ({\pi\over 2},{3\pi\over 2})$, as we can again see from
inspection of figure (2).
\vskip.5truecm\par\noindent
Finally we note that actually the  complex pair solutions come in pairs.  This
is because of the pairing of $\sqrt\omega$ for a particular $m$ odd with
$1\over\sqrt\omega$ for the corresponding $m$ even.  These yield identical
equations
(21) and (33).  The corresonding pairs of complex pair solutions are, however,
not
the quartets to which we had referred earlier.  Interestingly enough though,
clearly both pairs drop out simultaneously when the critical condition equation
(25)
is staisfied.  Thus actually two complex pair solutions of the ``string
hypothesis"
drop out simultaneously and become two real pair solutions.
\vskip.5truecm\par\noindent
We present below in somewhat less detail, the case when $N$ even.  First of
all,
we have $(\sqrt\omega )^N=(-1)^m$, for either choice for the square root,
$\sqrt\omega = \pm e^{i{\pi m\over N}}$.  Thus, for the case $m$ odd, we get
the
condition  as above,
$$
\cos({\theta\over 2})={\sin ({(N-2)\alpha\over 2})\over\sin ({N\alpha\over
2})},
\eqno(35)
$$
but here $N=2n$.   Thus,
$$
\cos({\theta\over 2})={\sin ((n-1)\alpha )\over\sin (n\alpha )}.\eqno(36)
$$
The R.H.S. is a function which has zeros at $\alpha ={k\pi\over (n-1)}$,
$k=1,2,3,\cdots ,2n-3$,  excluding $k=n-1$, i.e. $2n-4$ zeros, and poles at
$\alpha ={k\pi\over n}$, $k=1,2,\cdots ,2n-1$, excluding $k=n$. The value at
$\alpha =0$ is ${n-1\over n}$. The situation is as before except at $\alpha
=\pi$, the value is $-{n-1\over n}$, not $-1$, and the function ``turns over".
The function is reflection symmetric about $\alpha =\pi$.  We therefore obtain
for
each $\omega$ at least $2n-4$ real roots (the number of zeros), but we get two
additional real roots if
$$
\big|\cos({\theta\over 2})\big| \ge {n-1\over n},\eqno(37)
$$
making a total of $2n-2=N-2$.  This condition
can be satisfied for ${\theta\over 2}$ near zero or near $\pi$.  We also of
course get the reflections of these about $\pi $ hence we can restrict
${\theta\over 2}\in (0,\pi)$.  At $N=22$, $m=3$ is allowed, and then at $N=62$,
$m=5$ is allowed.   The condition becomes $m<{2\over\pi}\sqrt N$ and
$|N-m|<{2\over\pi}\sqrt N$ for large $N$.  Thus for the case $N$ even,
but $m$ odd there can be two solutions which violate the string hypothesis
correpondence.  We also always have solutions of the original polynomial
corresponding to $\tilde X_1 =\pm 1$, ($N$ even) which gives totally $N$ roots,
in which case, there are no complex roots for values of $m$ satisfying
equation (37).  We notice that equation (37) is satisfied simultaneously by two
values, $m_1,m_2$ such that $m_1+m_2=N$.  (N.B. $m_i$ are odd, $N$ is even.)
Hence we again lose two complex pair solutions, giving two additional real pair
solutions.
\vskip.5truecm\par\noindent
For $m$ even, the equation for $\tilde X_1$
becomes $$ \eqalign{
0&=(\omega +1)\sqrt\omega^N\tilde X_1^N-2\omega\sqrt\omega^{N-1}
\tilde X_1^{(N-1)}-2\sqrt\omega\tilde X_1+(\omega +1)\cr
&=(\omega +1)\tilde X_1^N-2\sqrt\omega \tilde X_1^{(N-1)}-
2\sqrt\omega \tilde X_1+(\omega +1)\cr
&=\sqrt\omega\left((\sqrt\omega+{1\over\sqrt\omega})\tilde X_1^N-2\tilde
X_1^{N-1}-2\tilde X_1+(\sqrt\omega+{1\over\sqrt\omega})\right).}\eqno(38)
$$
Replacing $\tilde X_1=e^{i\alpha}$ yields
$$
\cos({\theta\over 2})e^{iN\alpha}-e^{i(N-1)\alpha}-e^{i\alpha}
+\cos({\theta\over 2}) =0, \eqno(39)
$$
(with $\sqrt\omega = e^{i{\theta\over 2}}$) which simplifies to
$$
\eqalign{
0&=e^{iN\alpha\over 2}\cos({\theta\over 2})(e^{iN\alpha\over 2}+
e^{-iN\alpha\over 2})- e^{i\alpha}e^{i(N-2)\alpha\over 2}
(e^{i(N-2)\alpha\over 2}+e^{-i(N-2)\alpha\over 2})\cr
&=e^{iN\alpha\over 2}2\left(\cos({\theta\over 2})\cos ({N\alpha\over 2})-
\cos ({(N-2)\alpha\over 2})\right),}\eqno(40)
$$
yielding
$$
\cos({\theta\over 2})={\cos ({(N-2)\alpha\over 2})\over\cos ({N\alpha\over
2})}.
\eqno(41)
$$
This equation is also easily studied graphically, see figure (3).  The R.H.S.
has
zeros at $\alpha = {k\pi\over (N-2)}$ for $k=1,3,5,\cdots ,2N-5$ and poles at
$\alpha = {k\pi\over N}$ for $k=1,3,5,\cdots ,2N-1$.  We can check that at
$\alpha =
0$ the function is $1$ and at $\alpha =\pi$ it is $-1$.
Graphically we get a root for each intersection of a horizontal line $y=
\cos({\theta\over 2})$ with the curve $y={\cos ({(N-2)\alpha\over2})
\over\cos({N\alpha\over 2})}$.
\centinsert{\pscaption
{\boxit{\psboxscaled{800}{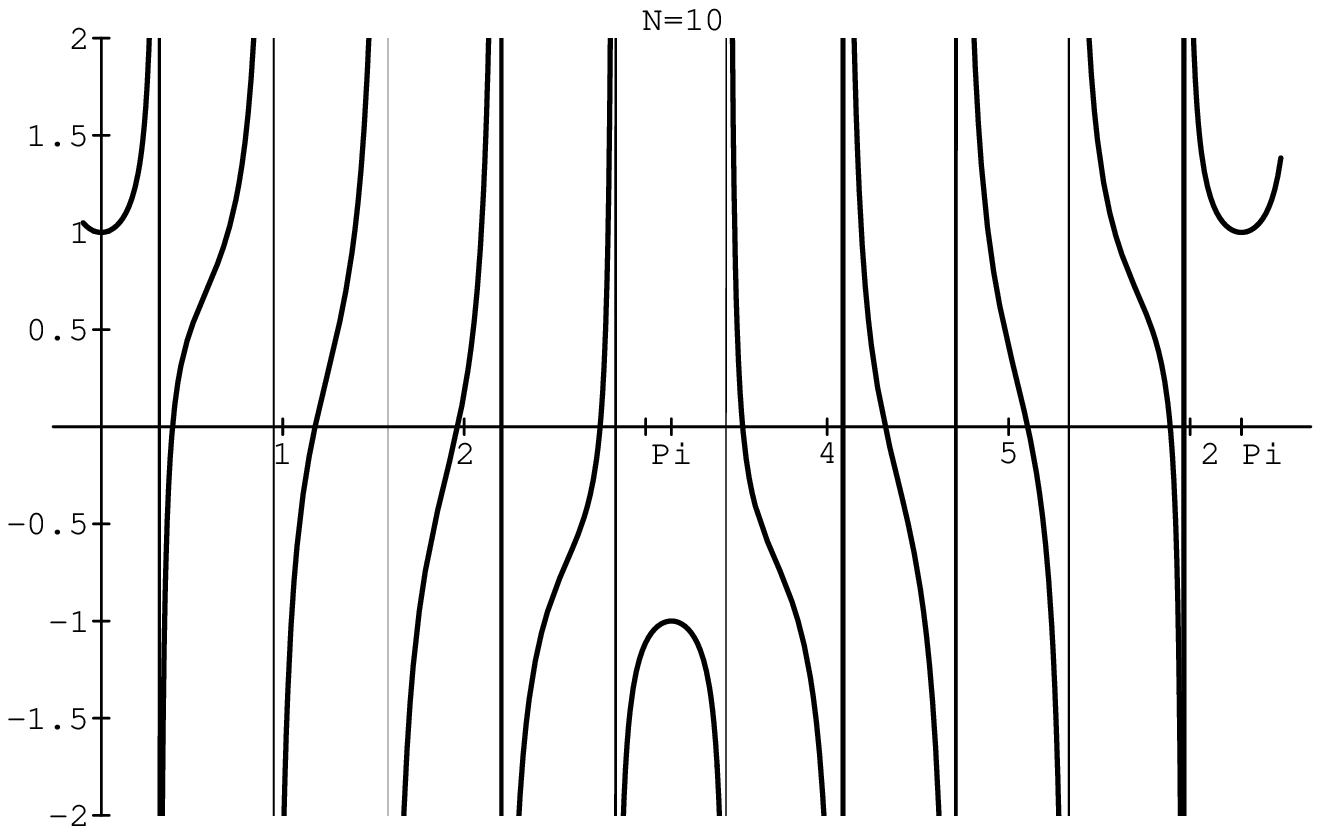}}}
{{\bf Figure 3}: $f(\alpha)={\cos ({(N-2)\alpha\over 2})\over\cos
({N\alpha\over
2})}$, $N=10$}}
\par\noindent
Counting the intersections it is evident from figure (3), that there is a root
for
each zero.  There are $N-2$ zeros, thus there are always only $N-2$ real
roots, $\tilde X_1=\pm 1$ are not roots in this case, in fact it is easy to see
that $\cos \left({N\alpha\over 2}\right)$ and  $\cos \left({(N-2)\alpha\over
2}\right)$ never vanish simultaneously.  Hence there are always two complex
roots and no new real roots which violate the string hypothesis
correspondence.
\vskip.5truecm\par\noindent
We summarize in the following way.  For $N$ sufficiently large, we
have $2\left({2\over\pi}\sqrt N\right)$ additional real pair solutions with
respect
to the prediction of the string hypothesis.  For $N$ odd, they are distributed
evenly between $m$ odd and $m$ even, however, they occur only near $\tilde
X_1=1$.
For $N$ even, they occur only for $m$ odd, however, now they are distributed
evenly
near $\tilde X_1=\pm 1$.  We give simple, exact expressions for the
critical values $N$, equations (25) and (37), which do not appear in
Reference (3).
\vskip.5truecm\par\noindent
Consider now the sector with $M$ spins flipped.  Here the equations (7),
rewritten in terms of the variables $X_i=e(\Lambda_i)$ are
$$
X_i^N=(-1)^{M-1}\prod_{l=1\atop l\ne i}^M({X_iX_l-2X_i+1\over
X_iX_l-2X_l+1}),\quad
i=1,\cdots ,M.\eqno(42)
$$
Multiplying the $M$ equations together gives
$$
(X_1X_2\cdots X_M)^N=1.\eqno(43)
$$
Thus
$$
X_M={\omega\over X_1X_2\cdots X_{M-1}}\eqno(44)
$$
with $\omega^N=1$.  Multiplying through with the denominator we get the coupled
polynomial system
$$
X_i\prod_{l=1\atop l\ne i}^M(X_iX_l-2X_l+1)+(-1)^M\prod_{l=1\atop l\ne
i}^M(X_iX_l-2X_i+1)=0\quad i=1,\cdots ,M.\eqno(45)
$$
Using equation (44) we can eliminate $X_M$ from the first $M-1$ equations and
multiplying through with the denominator yields the system of $M-1$ polynomial
equations
$$
\eqalign{
X_i&\prod_{l=1\atop l\ne
i}^{M-1}(X_iX_l-2X_l+1)((X_i-2)\omega +X_1X_2\cdots
X_{M-1})\cr
&+ (-1)^M\prod_{l=1\atop l\ne i}^{M-1}(X_iX_l-2X_i+1)(X_i\omega
-(2X_i-1)X_1X_2\cdots X_{M-1})=0\cr
&i=1,\cdots ,M-1.}\eqno(46)
$$
Each equation is just a permutation of the variables in any other one.
Removing
an overall factor of $X_i$, the $i$th equation is of order $N+M-2$ in $X_i$,
but
only quadratic in all the other variables.
\vskip.5truecm\par\noindent
Specializing to $M=3$ we find the system
$$
\eqalign{
0=X_1^N(X_1X_2-2X_2+1)&((X_1-2)\omega +X_1X_2)\cr
&-(X_1X_2-2X_1+1)(X_1\omega
-2X_1^2+X_1X_2)\cr
0=X_2^N(X_1X_2-2X_1+1)&((X_2-2)\omega +X_1X_2)\cr
&-(X_1X_2-2X_2+1)(X_2\omega -2X_2^2+X_1X_2).}\eqno(47)
$$
Simplifying we get
$$
\eqalign{
0&=X_2^2(X_1^{N+1}-2X_1^N+2X_1^2-X_1)+X_2(\omega(X_1^{N+1}-4X_1^N+4X_1^{N-1}
-X_1)\cr
&+(X_1^N-4X_1^2+4X_1-1))+\omega(X_1^N-2X_1^{N-1}+2X_1-1)\cr
X_1&\leftrightarrow X_2.}\eqno(48a,b)
$$
There are two ways to proceed, both give the same result up to trivial factors.
The simplest and straightforward method is to solve the quadratic equation
(48a)
for $X_2$  $$
A(X_1)X_2^2+B(X_1)X_2+C(X_1)=0\eqno(49)
$$
i.e.
$$
X_2={-B(X_1)\pm\sqrt{B^2(X_1)-4A(X_1)C(X_1)}\over 2A(X_1)}\eqno(50)
$$
as an algebraic function of $X_1$.  Then replace this in the corresponding
equation (48b) for $X_1$
$$
A(X_2)X_1^2+B(X_2)X_1+C(X_2)=0.\eqno(51)
$$
Expanding the powers of $X_2$  we can isolate the odd powers of the square root
to one side of the equation, the other side then only involves rational
polynomial expressions in $X_1$.  Finally squaring both sides and multiplying
through by the denominator gives a single polynomial equation for $X_1$.
Generalizing from calculations with Mathematica for low $N$, we obtain the form
of the polynomial as $(X_1-1)^{N+4}X_1^{N-2}A(X_1)^{N+1}P_N(X_1)$,
where $P_N$ is of degree $N(N-1)+1$.
It is not illuminating to present the explicit form of $P_N$ for a few explicit
value of $N$, we do not have a closed form for it.  The roots of
the polynomial along with the equivalent of equation (44) should give rise to
all solutions of the Bethe ansatz equations in this sector.
\vskip.5truecm\par\noindent
The second way to proceed affords generalization to higher values of $M$,
although this is not straightforward.  We consider the quadratic equation in
$X_2$ as giving the first step of a recurrence relation
$$
X_2^2=-{B(X_1)\over A(X_1)}X_2-{C(X_1)\over A(X_1)}.\eqno(52)
$$
Then if $X_2^n=\alpha_nX_2+\beta_n$ we have
$$
\eqalign{
\alpha_nX_2+\beta_n=X_2^n&=X_2X_2^{n-1}=X_2(\alpha_{n-1}X_2+\beta_{n-1})\cr
&=(\alpha_{n-1}X_2^2+\beta_{n-1}X_2)\cr
&=(\alpha_{n-1}(-{B(X_1)\over A(X_1)}X_2-{C(X_1)\over
A(X_1)})+\beta_{n-1}X_2)\cr
&=(\alpha_{n-1}(-{B(X_1)\over A(X_1)})+\beta_{n-1})X_2+\alpha_{n-1}
(-{C(X_1)\over A(X_1)}).}\eqno(53)
$$
Thus
$$
\pmatrix{\alpha_n\cr\beta_n}={1\over A(X_1)}\pmatrix{-B(X_1)\quad A(X_1)\cr
-C(X_1)\quad\quad 0}\pmatrix{\alpha_{n-1}\cr\beta_{n-1}}={\cal M}
\pmatrix{\alpha_{n-1}\cr\beta_{n-1}},\eqno(54)
$$
which is trivial to solve as
$$
\pmatrix{\alpha_n\cr\beta_n}={\cal M}^{n-2}\pmatrix{-{B(X_1)\over A(X_1)}\cr
-{C(X_1)\over A(X_1)}}.\eqno(55)
$$
Then replacing in equation (51) for $X_2^n$ we obtain a linear equation in
$X_2$
that is a rational polynomial expression in $X_1$ and we solve for $X_2$ as a
rational polynomial in $X_1$.  Replacing this back in equation (49) and
multiplying through by the denominator again gives a single polynomial equation
in $X_1$. We obtain from Mathematica the same polynomial as the above, modulo
some trivial factors, $(X_1-1)^{N+1}X_1^NA(X_1)^NP_N(X_1)$. However,
since we have no closed form expression for $P_N$, whose degree is of order
$N^2$,  it is actually quite
unfeasible to go much beyond $N=20$.  We are presently engaged in numerical and
analytical analyses of the roots of this polynomial to see if there are any new
violations of the string hypothesis.
\vskip.5truecm\par\noindent
We find it is a dramatic simplification to deal with even, coupled polynomial
equations (46) than the original transcendental equations.  We hope to extend
our
analysis to other models where the Bethe ansatz has proven useful.
\vskip.5truecm\par\noindent
We thank T. Gisiger, Y. Saint-Aubin, R. MacKenzie and V. Spiridonov for useful
discussions.  This work supported in part by NSERC of Canada, FCAR du Qu\'ebec
and FOM of the Netherlands.  We also thank Gebhard Gr\"ubl and the Institut
f\"ur
Theoretische Physik, Innsbruck, Austria for hospitality, where some of this
work
was done.
\vfill
\centerline{\bf References}
\vskip.5truecm\par\noindent
$^*$    address after November 1st, 1992, Sonnmatt 8, B\"ach, Switzerland.
\par\noindent
$^\dagger$    permanent address
\par\noindent
1.)  L. Faddeev and Takhtadjan, Cargese lectures,1987.
\par\noindent
2.)  H. Bethe, Z. Phys. 71, (1931)205.
\par\noindent
3.)  F.H.L. Essler, V.E. Korepin and K. Schoutens, J. Phys.A25,4115,(1992).
\par\noindent
4.) M. Takahashi, Prog. Theo. Phys.46, (1971) 401.
\vfill
\end